\documentclass[showpacs,twocolumn,aps,prb]{revtex4}
\usepackage{mathrsfs}
\usepackage{amsmath}
\usepackage{amsfonts}
\usepackage{amssymb}
\usepackage{amsthm}
\usepackage{graphicx}
\usepackage{color}
\usepackage{epsfig}
\usepackage{cancel}
\usepackage{bm}
\usepackage{sidecap}
\usepackage{epstopdf}

\providecommand{\U}[1]{\protect\rule{.1in}{.1in}}
\usepackage{ulem}
\begin{document}
%----------------------------------------------------------------
\title{Magnonic Weyl semimetal in pyrochlore ferromagnets}
\author{Ying Su$^{1,2}$}
\author{X. S. Wang$^{1,2}$}
\author{X. R. Wang$^{1,2}$}
\email[corresponding author: ]{phxwan@ust.hk}
\affiliation{$^{1}$Physics Department, The Hong Kong University of
Science and Technology, Clear Water Bay, Kowloon, Hong Kong}
\affiliation{$^{2}$HKUST Shenzhen Research Institute, Shenzhen 518057,
China}
\date{\today}
%------------------------------------------------------------------
\begin{abstract}
{Topological states of matter have been a subject of intensive 
studies in recent years because of their exotic properties such 
as the topologically protected edge and surface states. 
The initial studies were exclusively for electron systems. 
It is now known that topological states can also exist for other 
particles. Indeed, topologically protected edge states have already 
been found for phonons and photons. 
In spite of active searching for topological states in many fields,
the studies in magnetism are relatively rare although topological 
states are apparently important and useful in magnonics.
Here we show that the pyrochlore ferromagnets with the 
Dzyaloshinskii-Moriya interaction are intrinsic magnonic Weyl semimetals.
Similar to the electronic Weyl semimetals, the magnon bands in 
a magnonic Weyl semimetal are nontrivially crossing in pairs at 
special points (called Weyl nodes) in momentum space.  
The equal energy contour around the Weyl nodes gives rise to the 
Fermi arcs on sample surfaces due to the topologically protected 
surface states between each pair of Weyl nodes. 
Additional Weyl nodes and Fermi arcs can be generated in lower energy 
magnon bands when an anisotropic exchange interaction is introduced.}
\end{abstract}

%\pacs{73.43.Nq, 72.15.Rn, 72.25.-b, 85.75.-d}
\vspace{-0.2in}
\maketitle
%------------------------------------------------------------------

\section{Introduction}

Magnetic materials are highly correlated spin systems that do not
respect time-reversal symmetry. Their static states, such as 
domains, domain walls, and skyrmions, are the energy minimum spatial 
configurations of magnetization (vector order parameter) \cite{DW}.
The excitations of magnetic materials are spin waves whose quanta
are magnons of spin-1 particles. Like electrons, magnons can carry,
process and transmit information besides being a control knob
of magnetization dynamics \cite{yanpeng,hubin,xiansi}.
In fact, magnonics \cite{M1,M2,M3,M4,M5,M6,M7,M8} is a very 
active research field because of low energy consumption of magnonic 
devices and possible long spin coherence length \cite{D1,D2,D3}. 
One important issue in magnonics is the efficient transportation of magnons.
Magnon (spin wave) flux normally decays fast during its propagation
because it is difficult to confine magnons in the space.
Finding materials or structures that can confine the motion of magnons
in a restricted region under topological protection should open doors 
to new functional devices.
Thus, the realization of topological states of matter in magnetic
systems should be highly desirable \cite{TMI1,TMI2,WM}.

In this work, we show that the pyrochlore ferromagnet Lu$_2$V$_2$O$_7$, 
which was recently shown to exhibit magnon Hall effect \cite{MH}, 
is an intrinsic magnonic Weyl semimetal (MWS). 
%under an external magnetic field along [111] direction. 
Two adjacent magnon bands in a MWS nontrivially cross each other at some 
special points called Weyl nodes (WNs) in momentum space. 
The WNs are monopoles of Berry curvature and are characterized 
by integer topological charges. Because the net topological charges 
in the entire Brillouin zone (BZ) must be zero, the WNs must appear 
in pairs with opposite topological charges \cite{PLB}.
Like the electronic Weyl semimetal, the MWS has topologically 
protected chiral surface states between each pair of WNs on the 
sample surfaces \cite{EWS1,EWS2,EWS3}. 
%The equal energy contour of these surface states form arcs (called 
%Fermi arc), and the number of Fermi arcs between a pair of WNs 
%equals to the number of paired WNs. 
%topological charges carried by one of them. 
The equal
energy contour of these surface states form arcs (called
Fermi arc), and 
the number of Fermi arcs between two paired WNs equals to the number of topological charges carried by one of them.
Moreover, additional WNs and topologically protected surface states 
can appear in lower energy magnon bands when anisotropic exchange 
interaction, possibly induced by either doping or strain along the 
[111] direction, is introduced.   \\

\begin{figure*}
  \begin{center}% Requires \usepackage{graphicx}
  \includegraphics[width=15 cm]{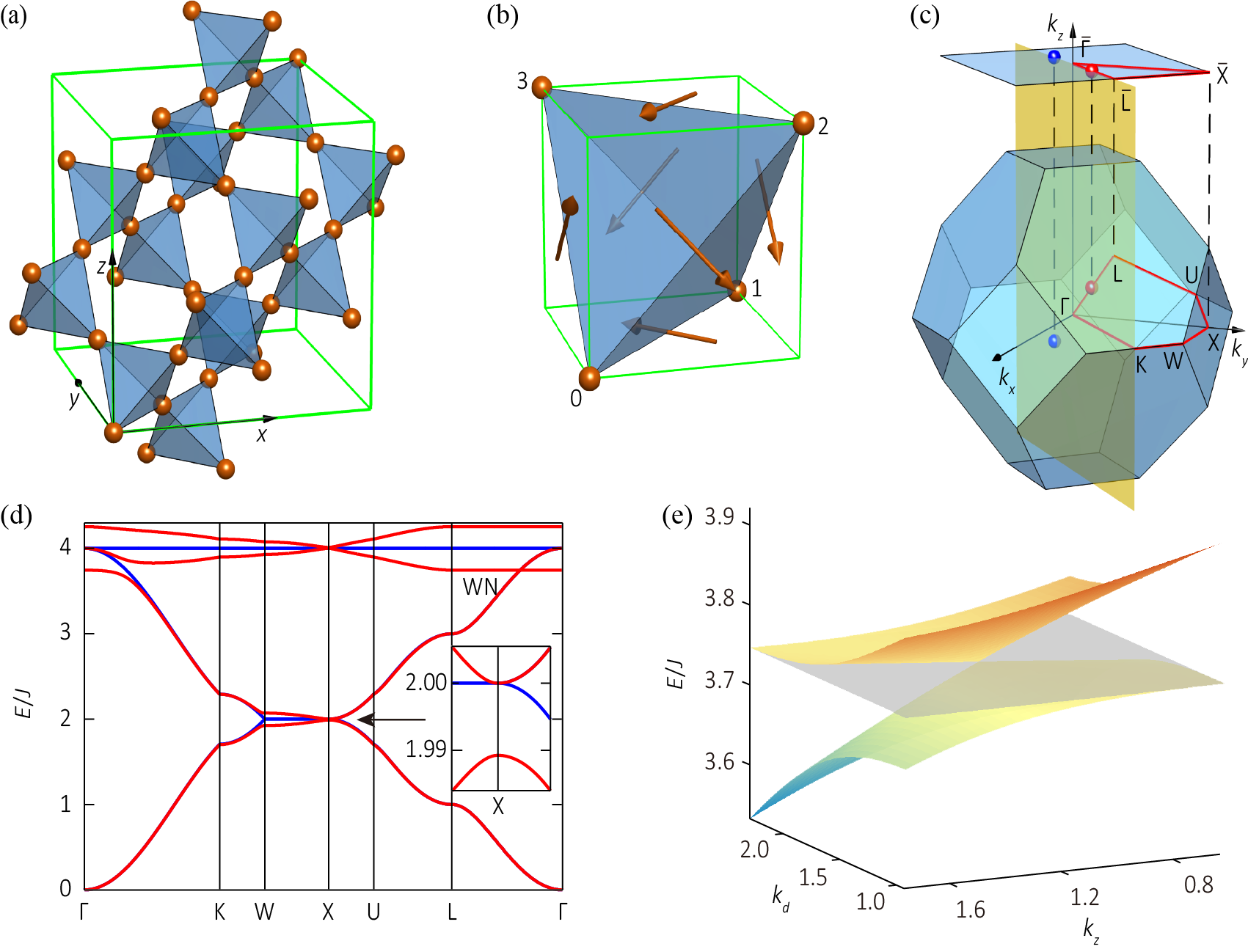}
  \end{center}
  \vspace{0in}
\caption{({a}) The pyrochlore structure of four interpenetrating 
face-centered cubic lattices with corner-sharing tetrahedrons.
The orange dots denote the V$^{4+}$ ions. ({b}) The DMI vector 
configuration of a tetrahedron lying in a corner-sharing cubic. 
The arrows represent the DMI vectors $\textbf{D}_{ij}$ perpendicular 
to the bonds of the tetrahedron and parallel to the surfaces of the cubic. 
({c}) The first bulk Brillouin zone (BZ) and the first (001) surface 
BZ of the pyrochlore lattice. 
%The three-dimensional first BZ of the pyrochlore lattice and its projection onto 
%the two-dimensional first BZ of a slab perpendicular to the [001] direction, where 
The projection of the high symmetry points of the bulk BZ onto the surface BZ  are denoted by the 
%The high symmetry points on the (001) surface BZ  are denoted by the 
barred symbols. The red and blue dots schematically represent the pair 
of WNs with opposite topological charges between $E_2$ and $E_3$ bands. 
({d}) The magnon energy spectrum along the high symmetry path shown 
in ({c}) with DMI (red curves) and without DMI (blue curves). 
({e}) The magnon dispersion around the WN shown in ({d}) in the 
$k_z$-$k_d$ plane which is represented by the yellow plane in ({c}).}
  \label{fig1}
  \vspace{-0.2in}
\end{figure*}

\section{Results}

\subsection{The effective spin model of Lu$_2$V$_2$O$_7$}

Lu$_2$V$_2$O$_7$ is an intrinsic ferromagnetic Mott-insulator in which 
each vanadium ion V$^{4+}$ carries spin $S=1/2$ \cite{MH,FM}. 
The magnetic properties of the material come purely from the vanadium ions 
that form a pyrochlore lattice consisting of four interpenetrating 
face-centered cubic (FCC) lattices with corner-sharing tetrahedrons, as 
shown in Fig. \ref{fig1}a. The primitive vectors are $\textbf{a}_1=(1,1,0)/2$, 
$\textbf{a}_2=(1,0,1)/2$, and $\textbf{a}_3=(0,1,1)/2$, where the FCC 
lattice constant is set to unit. Three of the four FCC lattices are shifted  
by $\textbf{a}_1/2$, $\textbf{a}_2/2$, and $\textbf{a}_3/2$, respectively. 
In each unit cell, there are four V$^{4+}$ ions as shown in Fig. \ref{fig1}b. 
Under an external magnetic field, the magnetic properties of the material 
is well described by a simple Heisenberg Hamiltonian with the Dzyaloshinskii-Moriya
interaction (DMI) \cite{MH}. 
The effective spin Hamiltonian reads  
\begin{equation}
H=-\sum_{\langle ij \rangle} J_{ij}\textbf{S}_i \cdot \textbf{S}_j
+ \sum_{\langle ij \rangle} \textbf{D}_{ij}\cdot \left(  \textbf{S}_i
\times \textbf{S}_j \right)  - g\mu_B \sum_i \textbf{h}\cdot \textbf{S}_i,
\label{H1}
\end{equation}
where $\langle ij \rangle$ denotes the nearest neighbor (NN) sites and 
$\textbf{S}_i$ is the spin of the V$^{4+}$ ion at site $i$. $\textbf{h}$ is
the external magnetic field applied along the [111] direction in this study. The first term describes the NN exchange interaction with strengths $J_{ij}$. The second term represents the DMI with the DMI vectors $\textbf{D}_{ij}$. The last term is the Zeeman interaction.

As it was explained in ref. \onlinecite{MH}, the material is a 
collinear ferromagnet in spite of the DMI because the summation 
of six DMI vectors adjacent to each lattice site is zero. Under the 
Holstein-Primakoff transformation and by using the Bloch theorem, 
the Hamiltonian (\ref{H1}) is block diagonalized in momentum space as 
$H=\sum_\textbf{k}b_\textbf{k}^\dagger \mathcal{H}(\textbf{k})b_\textbf{k}+E_0$,
where $b_\textbf{k}^\dagger=\left( b_{\textbf{k},0}^\dagger,b_{\textbf{k},1}
^\dagger,b_{\textbf{k},2}^\dagger,b_{\textbf{k},3}^\dagger \right)$ and
$b_\textbf{k}$ are the creation and annihilation operators of magnons (see Methods). 
The four components correspond to the four different FCC sublattices. 
$E_0=-Ng\mu_B|\textbf{h}|/
2 - \sum_i\sum_{j\in\langle ij\rangle} J_{ij}/8$ is the energy of zero magnon state (vacuum), where $N$ is the total number of lattice sites and ${j\in\langle ij\rangle}$ 
denotes $j$ as the NN site of $i$ (see Methods). One can set $E_0$ to zero by 
choosing a proper energy reference.
For a given $\textbf{k}$, $\mathcal{H}(\textbf{k})$ is a $4\times 4$ matrix
\begin{equation}
\begin{split}
&\mathcal{H}(\textbf{k}) = \left( \begin{array}{cccc}
3J & JA_1(\textbf{k}) & JA_2(\textbf{k}) & JA_3(\textbf{k}) \\
JA_1(\textbf{k}) & 3J & J_-A_{12}(\textbf{k}) & J_+A_{13}(\textbf{k}) \\
JA_2(\textbf{k}) & J_+A_{12}(\textbf{k}) & 3J & J_-A_{23}(\textbf{k})\\
JA_3(\textbf{k}) & J_-A_{13}(\textbf{k})  &
J_+A_{23}(\textbf{k}) & 3J
\end{array}\right),
\end{split}
\label{H2}
\end{equation}
where $A_\alpha(\textbf{k})=-\cos(\textbf{a}_\alpha\cdot\textbf{k}/2)$,
$A_{\alpha\beta}(\textbf{k})=-\cos[(\textbf{a}_\alpha-\textbf{a}_\beta)\cdot\textbf{k}/2]$, 
and $J_{ij}=J$ for isotropic exchange interaction.  
The strength of the DMI is a constant $D=|\textbf{D}_{ij}|$. 
Because only the components of $\textbf{D}_{ij}$ parallel to the external 
magnetic field contribute to Hamiltonian (\ref{H1}) (see Methods),
$J_\pm=J\pm i\sqrt{2}D/\sqrt{3}$
for $\textbf{h}$ along the [111] direction \cite{TMI1,MH}.
The magnitude of magnetic field is set as $|\textbf{h}|=0^+$ for simplicity (and without loss of generality) since the Zeeman interaction only shifts the magnon dispersion relation and does not affect the topological properties. 
In the absence of DMI $D=0$, the magnon spectrum contains two degenerate 
flat bands $E_i(\textbf{k})=4J$ ($i=1,2$) and two dispersive bands 
$E_i(\textbf{k})=2J\pm J\sqrt{1+F(\textbf{k})}$ ($i=3,4$), where $F(\textbf
{k})=\cos(k_x/2)\cos(k_y/2)+\cos(k_x/2)\cos(k_z/2)+\cos(k_y/2)\cos(k_z/2)$. 
The magnon dispersion relation along the high symmetry path $\Gamma$-K-W-X-U-L-$
\Gamma$ (see Fig. \ref{fig1}c) is shown in Fig. \ref{fig1}d for $D=0$ 
(blue curves) and for the experimental value $D=0.18J$ \cite{DM2} (red curves). 
In comparison with the case of $D=0$, the flat bands become dispersive and 
band gaps are opened.\\

\subsection{Identification of Weyl nodes and Fermi arcs}

Interestingly, a pair of WNs appears on the high symmetry line L-$\Gamma$-L as
shown in Figs \ref{fig1}c and \ref{fig1}d. Two magnon bands of $E_2$ and 
$E_3$ linearly cross each other, giving rise to a MWS behavior. 
Along the L-$\Gamma$-L line, where $\textbf{k}=k_1(1,1,1)$, two magnon bands are 
flat with $E_i(\textbf{k})=4J\pm \sqrt{2}D$ ($i=1,2$), and the other two bands 
are dispersive with $E_i(\textbf{k})=2J\pm J\sqrt{2.5+1.5\cos k_1}$ ($i=3,4$).
For $D=0$, $E_3$ touches $E_1$ and $E_2$ at the $\Gamma$ point. 
For modest $D> 0$, $E_1$ and $E_2$ are split into two nondegenerate flat bands, 
and $E_2$ and $E_3$ cross at a pair of WNs at 
\begin{equation}
k_1= \pm \cos^{-1}\left[ \frac{2(2J-\sqrt{2}D)^2-5J^2}{3J^2} \right].
\end{equation}
$k_1=\pm 1.198$ for $D=0.18J$. Moreover, the flatness of $E_2$ along the 
L-$\Gamma$-L line means the magnon group velocity near the 
WNs vanishes along the [111] direction. According to a recent classification \cite{WS2}, 
this corresponds to the transition state from type-I to type-II 
Weyl semimetals with vanishing group velocity only in one direction. 
To visualize the magnon dispersion with vanishing group velocity along the 
[111] direction, we plot the magnon bands of $E_2$ and $E_3$ near one WN in a 
vertical plane (represented by the yellow plane in Fig. \ref{fig1}c) parallel to 
both $k_z$ direction and the diagonal $k_x$-$k_y$ direction (termed as $k_d$). 
Obviously, the L-$\Gamma$-L line lies in the $k_z$-$k_d$ plane, and $E_2$ band 
and $E_3$ band linearly cross each other at the WN of $\textbf{k}=1.198(1,1,1)$ 
as shown in Fig. \ref{fig1}e. 

\begin{figure}
  \begin{center}% Requires \usepackage{graphicx}
  \includegraphics[width=8.5 cm]{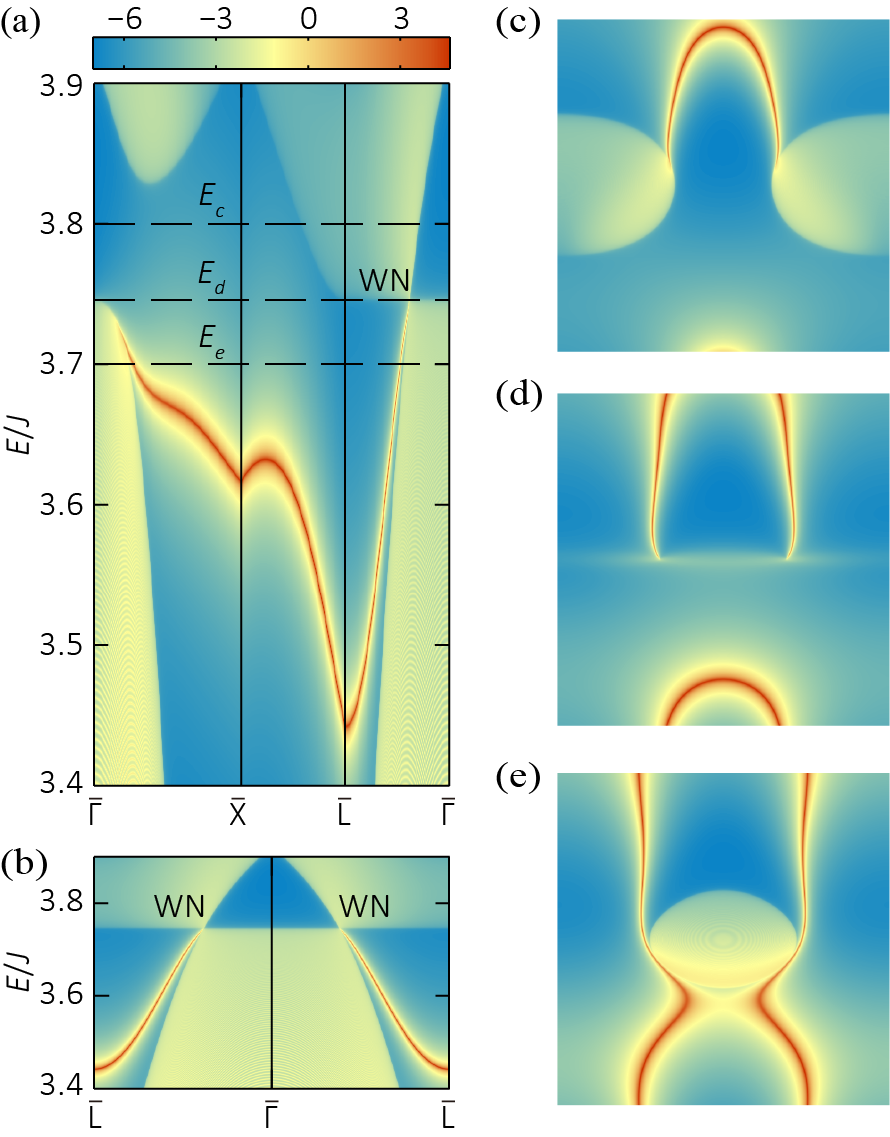}
  \end{center}
  \vspace{0in}
\caption{({a}) 
The density plot of magnon spectral function on the top (001) 
surface along the $\overline{\Gamma}$-$\overline{\text{X}}$-$\overline
{\text{L}}$-$\overline{\Gamma}$ path.
({b}) The density plot of magnon spectral function on the top (001) surface 
along the $\overline{\text{L}}$-$\overline{\Gamma}$-$\overline{\text{L}}$ path.
({c})-({e}) The density plot of magnon spectral function
on the top (001) surface in the first BZ for fixed energies of 
$E_c$, $E_d$, and $E_e$ denoted in ({a}).} 
  \label{fig2}
  \vspace{-0.2in}
\end{figure}

Similar to the electronic Weyl semimetal, one fingerprint of the MWS 
is the Fermi arcs on the sample surfaces. In order to illustrate this feature, 
we consider a slab whose surfaces are perpendicular to the [001] direction. 
The first BZ of the (001) surface is shown in Fig. \ref{fig1}c, where the 
projection of the high symmetry points of the first bulk BZ onto the first 
surface BZ are denoted by the barred symbols. The pair of WNs are 
schematically represented by the red and blue dots (indicating they carry 
opposite topological charges) in Fig. \ref{fig1}c.
The density plot of magnon spectral function on the top surface along the high 
symmetry path of $\overline{\Gamma}$-$\overline{\text{X}}$-$\overline{\text{L}}$-$
\overline{\Gamma}$ is shown in Fig. \ref{fig2}a where one WN can be identified. 
%The L-$\Gamma$ path is projected to the $\overline{\text{L}}$-$\overline
%{\Gamma}$ path on which one of the pair of WNs are identified in two 
%dimensional momentum space as shown in Fig. \ref{fig2}a.  
The topologically protected surface states with high density on the top 
surface are represented by red color. On the path of $\overline{\text{L}}
$-$\overline{\Gamma}$-$\overline{\text{L}}$ where both of the two WNs 
lies in, the density plot of magnon spectral function on the top surface 
is shown in Fig.~\ref{fig2}b. Apparently, the pair of WNs are connected 
by surface states. For fixed energies of $E_c$, $E_d$, and $E_e$ 
around the WNs (see Fig.~\ref{fig2}a), the corresponding density 
plot of magnon spectral function on the top surface 
in the first BZ are shown in Figs \ref{fig2}c-\ref{fig2}e, respectively. 
The Fermi arc formed by topologically protected surface states on the 
top surface is clearly displayed. 
For $E_d=4J-\sqrt{2}D=3.745J$ through the WNs, the Fermi arc is 
terminated at the two WNs as shown in Fig. \ref{fig2}d. \\

\begin{figure*}
  \begin{center}% Requires \usepackage{graphicx}
  \includegraphics[width=17 cm]{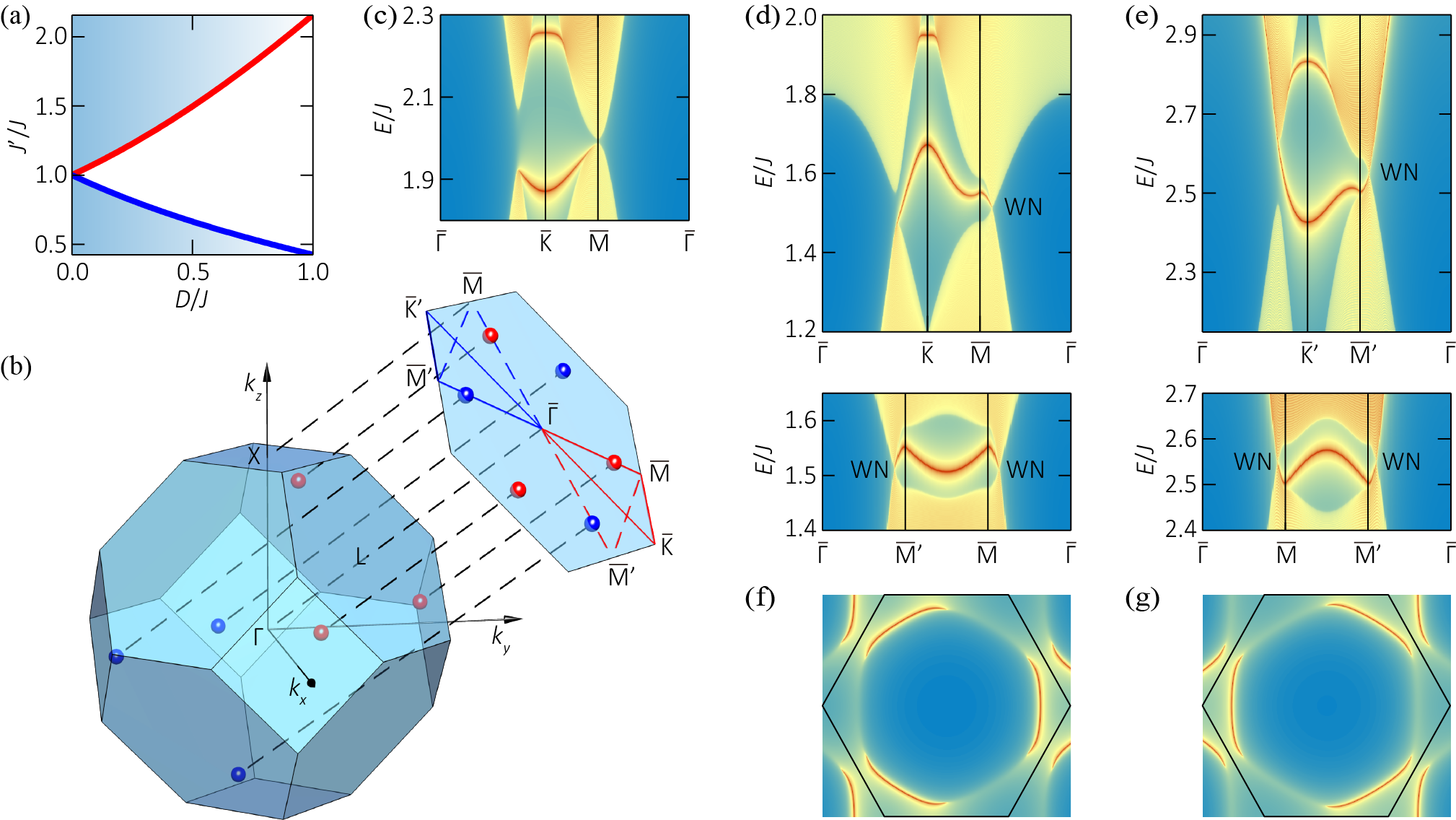}
  \end{center}
  \vspace{0in}
\caption{({a}) The phase diagram of the MWS, due to $E_3$ and $E_4$ bands, 
in the $D$-$J'$ plane. The MWS, together with WNs and Fermi arcs, exists in the shadowed regions. The white region between the two critical interlayer exchange interactions $J_\pm'(D)$ 
denoted by red and blue curves is the normal magnonic insulator 
without topologically protected surface states in the energy gap. 
({b}) The first bulk Brillouin zone (BZ) and the first (111) surface BZ of the 
pyrochlore lattice. The red and blue dots schematically represent 
the three pairs of WNs with opposite topological charges between the bands of 
$E_3$ and $E_4$. 
({c}) The density plot of magnon spectral function on the top surface 
for $J'=J$ along the path of $\overline{\Gamma}
$-$\overline{\text{K}}$-$\overline{\text{M}}$-$\overline{\Gamma}$. 
({d}) The density plot of magnon spectral function on the top surface 
for $J'=0.6J$ along the paths of $\overline{\Gamma}$-$\overline{\text{K}}$-$
\overline{\text{M}}$-$\overline{\Gamma}$ (upper panel) and $\overline{\Gamma}
$-$\overline{\text{M}}'$-$\overline{\text{M}}$-$\overline{\Gamma}$ (lower 
panel) that are presented by red solid and dash lines in ({b}). 
({e}) The density plot of magnon spectral function on the top surface 
for $J'=1.6J$ along the paths of $\overline{\Gamma}$-$\overline{\text{K}}'$-$
\overline{\text{M}}'$-$\overline{\Gamma}$ (upper panel) and $\overline
{\Gamma}$-$\overline{\text{M}}$-$\overline{\text{M}}'$-$\overline{\Gamma}$ 
(lower panel) that are presented by blue solid and dash lines in ({b}). 
({f})-({g}) The Fermi arcs on the top (111) surface for  energies through the WNs for $J'=0.6J<J'_-$ ({f}) 
and $J'=1.6J>J'_+$ ({g}). 
The black hexagon encloses the first BZ.}
  \label{fig3}
  \vspace{-0.2in}
\end{figure*}

\subsection{Anisotropic exchange interaction}

We have shown that the pyrochlore ferromagnet Lu$_2$V$_2$O$_7$ is 
an intrinsic MWS. We would like to show now that more pairs of WNs and 
topologically protected surface states can come from the lower energy magnon bands of $E_3$ and 
$E_4$ in the presence of anisotropic exchange interaction. 
The pyrochlore lattice can be viewed as an alternative stack of kagome and 
triangular lattices along the [111] direction. In principle, the interlayer 
exchange interaction $J'$ differs from the intralayer exchange interaction $J$. 
$J'$ can be tuned by either doping \cite{TMI1} or strain \cite{DT}. 
The effective Hamiltonian with the anisotropic exchange interaction, under the 
same considerations as before, becomes  
\begin{equation}
\begin{split}
\mathcal{H}'(\textbf{k})=
\left( \begin{array}{cccc}
3J' & J'A_1(\textbf{k}) & J'A_2(\textbf{k}) & J'A_3(\textbf{k}) \\
J'A_1(\textbf{k})& 2J+J' & J_-A_{12}(\textbf{k})& J_+A_{13}(\textbf{k})\\
J'A_2(\textbf{k})&J_+A_{12}(\textbf{k})& 2J+J' & J_-A_{23}(\textbf{k})\\
J'A_3(\textbf{k}) & J_-A_{13}(\textbf{k})  &
J_+A_{23}(\textbf{k}) & 2J+J'
\end{array}\right).
\end{split}
\label{H3}
\end{equation} 
The anisotropic exchange interaction leads to different sublattice on-site 
potential because the one-site potential on each particular site is the sum 
of the exchange interaction strengths of all its NNs (see Methods). 

For the isotropic exchange interaction, the energy gap minimum between $E_3$ 
and $E_4$ bands is at the X point as shown in Fig.~\ref{fig1}d. 
The anisotropic exchange interaction can close the gap at the X point whenever 
$J'$ equals to two critical values
\begin{equation}
J'_\pm = \alpha \pm \sqrt{2\alpha^2 - 2\alpha J},
\end{equation}
where $\alpha = \sqrt{J^2+2D^2/3}$. The critical $J'_\pm$ as functions of the
DMI strength $D$ are plotted as red and blue curves in Fig.~\ref{fig3}a. 
These are the phase boundaries between the normal magnonic insulator (without 
topologically protected surface states in the gap) and the MWS from 
$E_3$ and $E_4$ bands. The phase diagram of the MWS from $E_3$ and 
$E_4$ bands in the $D$-$J'$ plane is shown in Fig.~\ref{fig3}a. 
%together of course with WNs and topologically protected surface states. 
In the shadowed regions of Fig.~\ref{fig3}a where $J'>J'_+$ or $J'_->J'>0$, 
$E_3$ and $E_4$ bands always cross at three pairs of WNs due to the three-fold 
rotation symmetry with respect to the L-$\Gamma$-L line (see Fig.~\ref
{fig3}b). For Lu$_2$V$_2$O$_7$, $J'_+=1.158J$  and $J'_-=0.863J$. 
In the limits of $J'\rightarrow \infty$ and 0, all these WNs will merge at the 
L point. The fact that the trivial region represented by white color shrinks 
as $D$ decreases means that weak DMI is favorable for the existence of WNs 
between $E_3$ and $E_4$ bands since only weak anisotropy (small difference 
between the interlayer and intralayer exchange interactions) is required.  
These results are applicable to other pyrochlore ferromagnets. \\

\subsection{Additional Weyl nodes and Fermi arcs}

To visualize these additional WNs and topologically protected surface 
states existing in the MWS phase from $E_3$ and $E_4$ bands, the 
magnon spectral function of a slab with (111) surfaces is calculated. 
The density plot of magnon spectral function on the top surface for $J'=J$ 
along the high symmetry path $\overline{\Gamma}$-$\overline{\text{K}}
$-$\overline{\text{M}}$-$\overline{\Gamma}$ (marked by red solid 
lines in Fig. \ref{fig3}b) is shown in Fig. \ref{fig3}c. 
The energy gap minimum appears at the $\overline{\text{M}}$ point to which the X point
is projected. 
As the interlayer exchange interaction decreases to $J'=0.6J<J'_-$, three 
pairs of WNs are created from the linear crossing of $E_3$ and $E_4$ bands. 
The density plot of magnon spectral function on the top surface 
along various paths is shown in Fig.~\ref{fig3}d. 
Along the path of $\overline{\Gamma}$-$\overline{\text{K}}$-$
\overline{\text{M}}$-$\overline{\Gamma}$, a WN is identified on the 
$\overline{\text{M}}$-$\overline{\Gamma}$ segment. 
The topologically protected surface states is clearly visible within the 
energy gap with one end terminated at the WN. Along the path of $\overline
{\Gamma}$-$\overline{\text{M}}'$-$\overline{\text{M}}$-$\overline{\Gamma}$ 
(represented by red dash lines in Fig.~\ref{fig3}b), a pair of WNs is 
connected by the surface  states. Similar results for $J'=1.6J>J'_+$ are 
shown in Fig.~\ref{fig3}e along the $\overline{\Gamma}$-$\overline{\text
{K}}'$-$\overline{\text{M}}'$-$\overline{\Gamma}$ and $\overline{\Gamma}
$-$\overline{\text{M}}$-$\overline{\text{M}}'$-$\overline{\Gamma}$ paths 
(marked by blue solid and dash lines, respectively, in Fig.~\ref{fig3}b). 
In order to detect the Fermi arc feature, we fix the energy through 
the WNs for the two different interlayer exchange interaction strengths. 
The density plot of magnon spectral function on the top surface in the 
two-dimensional momentum space is shown in Figs \ref{fig3}f and 
\ref{fig3}g where the black hexagon encloses the first surface BZ. 
Apparently, the topologically protected surface states form three 
Fermi arcs of the three pairs of WNs. 
%The ends of the Fermi arcs show the position of WNs.

The pair of WNs from $E_2$ and $E_3$ bands on the L-$\Gamma$-L 
line can remain for the anisotropic exchange interaction.
Since the magnon dispersions on the L-$\Gamma$-L line with $\textbf{k}
=k_1(1,1,1)$ are $E_2(\textbf{k})=3J+J'-\sqrt{2}D$ and $E_3(\textbf{k})
=2J'+ J'\sqrt{2.5+1.5\cos k_1}$ in the present case, 
$E_2$ and $E_3$ bands cross at a pair of WNs at
%between $E_{2,3}(\textbf{k})$ are shifted to
\begin{equation}
k_1 = \pm \cos^{-1}\left[ \frac{2(3J-J'-\sqrt{2}D)^2-5J'^2}{3J'^2} \right],
\end{equation}
as long as $J-\sqrt{2}D/3<J'<3J/2-D/\sqrt{2}$.
Because the $E_2$ band is flat along the L-$\Gamma$-L line, the magnon 
group velocity around the pair of WNs vanishes along the [111] direction. \\

\section{Discussion}

\noindent The pyrochlore ferromagnet Lu$_2$V$_2$O$_7$ is an intrinsic 
topological material (called MWS) in the sense that two adjacent magnon 
bulk bands of $E_2$ and $E_3$ linearly cross each other at a special 
pair of points (called WNs) on the L-$\Gamma$-L line in momentum space. 
The distance between the paired WNs is determined by the strength of DMI. 
Similar to its electronic counterpart, the MWS has topologically 
protected chiral surface states whose equal energy contour yields 
the Fermi arc that connects the pair of WNs on the sample surfaces. 
By introducing different interlayer and intralayer exchange interaction strengths
through either doping or strain along the [111] direction, three 
additional pairs of WNs can be generated from the lower energy magnon bands of $E_3$ and $E_4$. 
On the surfaces of a slab perpendicular to the [111] direction, the three pairs 
of WNs are connected by three Fermi arcs in two-dimensional momentum space. 
Furthermore, the pair of WNs between $E_2$ and $E_3$ bands can remain on the 
L-$\Gamma$-L line whose distance is determined by both the DMI and 
interlayer exchange interaction. These results are applicable to other 
collinear pyrochlore ferromagnets with anisotropic exchange interaction.

The MWS featured by WNs and Fermi arcs can be detected by inelastic neutron 
scattering which has been used to probe the magnon bands of a topological magnon 
insulator \cite{TMI2}. The topologically protected magnon surface states 
can also be probed by the spin-polarized scanning tunneling microscopy through the second-order derivative of tunneling 
current that contains the information of electron-magnon scattering \cite{SPSTM1}.  \\

\section{Methods}

\subsection{Holstein-Primakoff transformation}
 
In this transformation \cite{HPT}, the spin-1/2 operators 
are mapped to the magnon creation and annihilation operators as
\begin{equation}
S^+_i = \sqrt{1-n_i} b_i,\quad S^-_i= 
b_i^\dagger \sqrt{1-n_i},\quad n_i=b_i^\dagger b_i,
\end{equation}
where the ladder operators $S_i^\pm=S_i^l\pm iS_i^m$ are defined in the orthonormal 
coordinate $(l,m,n)$ with $n$ axis parallel to the external magnetic field. 
For the DMI, the local spin $\textbf{S}_i = \textbf{S}+ \delta\textbf{S}_i$ 
where $\textbf{S}=(0,0,1/2)$ and $\delta\textbf{S}_i = (S_i^l,S_i^m,0)$ 
in the linear approximation. Thus, the Hamiltonian of DMI is
\begin{equation}
\begin{split}
H_{\rm DMI}&=\sum_{\langle ij\rangle}\textbf{D}_{ij}\cdot\left(\textbf{S}
\times\delta\textbf{S}_j + \delta\textbf{S}_i\times\textbf{S}+\delta
\textbf{S}_i\times\delta\textbf{S}_j\right)  \\
&=\sum_{\langle ij\rangle}\textbf{D}_{ij}\cdot\left( 0,0,S_i^lS_j^m-S_i^mS_j^l 
\right) \\ &=\sum_{\langle ij\rangle}\frac{iD_{ij}^n}{2}\left(S_i^+S_j^- - S_i^-S_j^+\right),
\end{split}
\end{equation}
where $D^n_{ij}=\textbf{D}_{ij}\cdot\hat{n}$ and $\sum_{\langle ij\rangle}
\textbf{D}_{ij}\cdot\left(\textbf{S}\times\delta\textbf{S}_j + \delta
\textbf{S}_i\times\textbf{S}\right)=0$. Namely, the $\textbf{D}_{ij}$ with 
vanishing $n$ component does not  contribute to the Hamiltonian. 
Substitute these into the effective spin Hamiltonian (\ref{H1}), 
we get a tight-binding Hamiltonian of magnons as 
\begin{equation}
\begin{split}
H=&- \frac{1}{2}\sum_{\langle ij \rangle}\left[  (J_{ij}+iD_{ij}^n) 
b_i^\dagger b_j+ {\rm H.c.} \right] + E_0, \\
&+\sum_{i}\left(\sum_{j\in\langle ij\rangle} \frac{J_{ij}}{2} + 
g\mu_B|\textbf{h}| \right)b_i^\dagger b_i. \\
\end{split}
\label{H4}
\end{equation}
Here ${j\in\langle ij\rangle}$ 
denotes $j$ as the NN site of $i$, and $E_0=-{Ng\mu_B|\textbf{h}|}
/{2} - \sum_i\sum_{j\in\langle ij\rangle} {J_{ij}}/{8}$ can be set to 
zero by choosing a proper energy reference, where $N$ is the  total number of lattice sties. 
Moreover, the on-site potential of each lattice site is determined by the sum of  all 
adjacent NN exchange interaction strengths such that the anisotropic exchange 
interaction can generate different sublattice on-site potential 
as shown in equation (\ref{H3}).

According to the Bloch theorem, the Hamiltonian (\ref{H3}) 
is block diagonalized in the basis of Bloch states 
\begin{equation}
|\textbf{k},\alpha\rangle = \frac{1}{\sqrt{N/4}} 
\sum_i e^{i\textbf{k}\cdot\textbf{r}_i} |i,\alpha\rangle,
\end{equation}
where $\alpha=0,1,2,3$ denote four different sublattices shown in Fig. 
\ref{fig1}b and $\textbf{r}_i$ is the position of the $i$th unit cell. 
Thus, we obtain the Hamiltonian (\ref{H2}) and (\ref{H3}). \\

\subsection{Surface spectral function.} 
The spectral function of a specific layer is
\begin{equation}
A_l(\textbf{k},E)=-\frac{1}{\pi}\text{Im} \left[ \text{Tr} G_{ll}
(E,\textbf{k}) \right],
\end{equation} 
where $l$ is the layer index and 
$G_{ll}(E,\textbf{k})=\langle l |(E+i0^+-H)^{-1}|l\rangle$.
For the top surface with $l=1$, $G_{11}(E,\textbf{k})$ is obtained 
by the recursive Green's function method \cite{RGF1,RGF2}.

{\it Note added.} Upon completion of this work, we became
aware of ref. \onlinecite{WM2}, in which part of the results were obtained.\\

\section{Acknowledgments}

This work is supported by the NSF of China Grant (No. 11374249)
and Hong Kong RGC Grants (No. 163011151 and No. 605413).

\end{document}